\documentclass[aps, prb, twocolumn, showpacs, superscriptaddress]{revtex4-1}

\usepackage{amsmath,amssymb}
\usepackage{graphicx}
\usepackage{dcolumn}
\usepackage{bm}

\begin{document}

\title {Two-dimensional confinement of critical U(1) gauge field in the presence of Fermi surface}
\author{Yin Zhong}
\email{zhongy05@hotmail.com}
\affiliation{Center for Interdisciplinary Studies $\&$ Key Laboratory for
Magnetism and Magnetic Materials of the MoE, Lanzhou University, Lanzhou 730000, China}
\author{Ke Liu}
\affiliation{Institute of Theoretical Physics, Lanzhou University, Lanzhou 730000, China}
\author{Yong-Qiang Wang}
\affiliation{Institute of Theoretical Physics, Lanzhou University, Lanzhou 730000, China}
\author{Hong-Gang Luo}
\email{luohg@lzu.edu.cn}
\affiliation{Center for Interdisciplinary Studies $\&$ Key Laboratory for
Magnetism and Magnetic Materials of the MoE, Lanzhou University, Lanzhou 730000, China}
\affiliation{Beijing Computational Science Research Center, Beijing 100084, China}


\begin{abstract}
The compact U(1) gauge field occurs in many fractionalized descriptions of low dimensional quantum magnetism and heavy fermion systems. In the presence of Fermi surface a fundamental and interesting question about the gauge field is whether it is confined or not. In this paper we find that the U(1) gauge field with a possible positive anomalous dimension is confined in two spatial dimensions, even it is strongly coupled to abundant gapless fermionic excitations near the Fermi surface. This result means that some quantum spin liquids and/or fractionalized metallic states described in terms of this kind of gauge fields are unstable, thus they are not be appropriate candidates for ground states in such systems. The instability obtained shows some novel features of the quantum spin liquids and/or fractionalized metallic states, which have not been reported in the literature. The result could be useful for further study in the quantum spin liquids and many related slave-particle gauge theories in strongly correlated electron systems.
\end{abstract}

\maketitle

\section{Introduction} \label{intr}
Understanding the nature of elusive quantum criticality is one of central issues in modern condensed matter physics.\cite{Sachdev2011,Sachdev2003,Rosch,Sachdev2008,Gegenwart,Si} One route to attack this challenging problem is to utilize the concepts and methodology of fractionalization.\cite{Wen,Senthil2003,Senthil2004,Senthil3,Senthil4,
Florens,Lee2005,Pepin2005,Kim2006,Senthil2008,Kim2010,Senthil2010} Its basic idea is that near the putative quantum critical points many elementary degrees of freedom such as magnons and electrons are broken into more elementary quasiparticles such as spinon and holon due to wild quantum fluctuations. Generically, the related theories with these fractionalized excitations involve emergent U(1) and/or even SU(2) gauge fields.\cite{Wen} Since these gauge fields emerge from underlying microscopic lattice models, they are compact in nature, which allows for a topological defect called instanton (or monopole) describing tunneling events between energetically degenerated but topological inequivalent ground states of gauge fields.\cite{Wen,Kogut,Polyakov1975,Polyakov1977}

It is argued many years ago by Polyakov that the pure compact U(1) gauge field is always in the confined state at zero temperature due to the ineluctable proliferation (condensation) of monopoles in two space dimensions.\cite{Polyakov1975,Polyakov1977} However, it is delicate to answer whether the compact U(1) gauge fields are confined or not in the presence of Fermi surface. Many previous studies based on effective sine-Gordon action have given controversial conclusions on the confinement of compact gauge fields.\cite{Ioffe,Nagaosa,Ichinose,Herbut2003,Kim,Kaul}(It is noted that the confinement of U(1) gauge fields coupled to massless Dirac fermions (the so-called $cQED_{3}$) is still controversial as well.\cite{Herbut2002,Herbut2003b,Case,Hermele,Unsal} However, a recent numerical simulation seems suggest that only confined states exist in the physically interesting condition.\cite{Armour}) Interestingly, the starting points of these studies are same, both result from one-loop renormalization-group (RG) or equivalently from conventional random-phases approximation (RPA). The results of RG or RPA indicate that the transverse gauge field is damped (acquiring a dynamical critical exponent $z=3$) due to its coupling to gapless fermions near Fermi surface and has no anomalous dimension and mass term. The absence of mass term of the transverse gauge field results from gauge invariance while the vanishing of anomalous dimension is not protected by any underlying principles or symmetries.

Lately, a nonpertubative argument, which formulates low-energy modes near the Fermi surface in terms of an infinite number of 1+1D chiral fermions, suggests that the U(1) gauge fields are deconfined for any nonzero favors of fermions.\cite{Lee2008} However the effective action with chiral fermions only includes the dispersion of the fermions transverse to the Fermi surface and the curvature of the Fermi surface is ignored. Thus the decay and breakdown of quasiparticles cannot be easily captured.\cite{Sachdev2011} Recently, quantum phase transition of metals in two space dimensions have been reexamined, and interesting scaling properties are also proposed for the dynamics of a Fermi surface with the fermions coupled minimally to a noncompact U(1) gauge field.\cite{Metlitski} Interestingly, a universal scaling propagator for the transverse gauge field and a RG equation for gauge charges are derived from the corresponding Ward identities.\cite{Metlitski,Mross2010} In principle, these results admit a nonzero anomalous dimension for gauge fields, particularly, for the physically interesting case of a positive one though they have not found a nontrivial anomalous dimension of gauge fields up to the three-loop order. Therefore, in the presence of Fermi surface an interesting and fundamental question is raised, namely, \emph{whether the compact U(1) gauge field with a positive anomalous dimension in two space dimensions is confined or not}.

In this paper, we propose a possible answer to this question. With the help of scaling properties derived by Metlitski and Sachdev\cite{Metlitski} and an effective sine-Gordon-like action for dual magnetic monopole potentials, we obtain approximate RG equations for gauge charges (or magnetic charges) and the fugacity of monopoles. From the RG equations, we find that the fugacity flows to infinity under the scale transformation and thus the proliferation of monopoles is inevitable. Consequently, the compact U(1) gauge field has to be in the confined state though it is strongly coupled to gapless fermionic excitations around Fermi surface. Therefore, it is likely that some
descriptions of quantum spin liquid and/or fractionalized metallic states involving compact U(1) gauge fields and gapless fermions with Fermi surface in two spatial dimensions are inherently unstable to other possible symmetry-breaking states.\cite{Senthil2003,Senthil2004,Lee2005,Paul,Pepin,Kaul,Galitski,Moon} Although we focus on the case of U(1) gauge fields in the present work, our argument appears to have a direct generalization to the SU(2) case.

The remainder of the present paper is organized as follows. In Sec. \ref{sec2}, we briefly review basic scaling properties proposed by Metlitski and Sachdev for noncompact U(1) gauge fields in the presence of Fermi surface in two space dimensions. In Sec. \ref{sec3}, we begin with an effective action of the gauge field by using a scaling propagator of the transverse gauge field as an input. After a dual mapping, a sine-Gordon-like action for dual magnetic monopole potentials is obtained. Its corresponding RG equations for magnetic charges and fugacity are derived and analyzed in Sec. \ref{sec4}. We find that fugacity flows to infinity under scale transformation, thus the compact U(1) gauge field is in the confined state due to the condensation of monopoles since the condensation corresponds to the infinite fugacity in low-energy limit. Finally, a concise conclusion is devoted to Sec. \ref{sec5}.

\section{scaling without monopoles} \label{sec2}
In this section, we firstly consider the scaling properties of a noncompact U(1) gauge field, the noncompactness means that excitations of monopoles can be safely neglected in low-energy limit.

Most studies on confinement/deconfinement of compact U(1) gauge fields coupled to Fermi surface are based on some sine-Gordon-like effective actions obtained by integrating over gapless fermions on Fermi surface.\cite{Ioffe,Nagaosa,Ichinose,Herbut2003,Kim,Kaul} The dynamics of gauge fields is damped by particle-hole excitation of fermions around Fermi surface, thus
the gauge field has a damped propagator $D^{-1}(k,\omega)=k^{2}+\gamma\frac{|\omega|}{k}$ ( here and hereafter $\omega$ is the imaginary frequency), where $\gamma$ denotes Landau damping parameter.\cite{Altshuler,Wen} It is noted that the excitation spectrum of gauge bosons is modified to $\omega\sim k^{3}$ in comparing to the standard form $\omega\sim k$, which is correct only for pure noncompact U(1) gauge fields. Meanwhile, the fermions suffer from scattering due to the damped gauge bosons and they acquire a singular self-energy correction with the form $\Sigma(k,\omega)\propto-isgn(\omega)|\omega|^{\frac{3}{2}}$ in two space dimensions.\cite{Altshuler,Wen}

For a long time it was thought that the above RPA's result was the exact form.\cite{Chubukov} However, recent studies pointed out that it may not be true based on direct perturbative  calculations up to three-loop order and compelling physical arguments.\cite{Lee2009,Metlitski,Mross2010} In stead of a further delicate perturbative calculation, in terms of the Ward identities of the corresponding low-energy effective two-patch action, Metlitski and Sachdev\cite{Metlitski} have obtained many interesting scaling properties for the dynamics of a Fermi surface with the fermions coupled minimally to a noncompact U(1) gauge field. For our purpose, we are interested in the scaling relations for the propagator of the transverse gauge field and a RG equation for gauge charges, which read as follows:
\begin{eqnarray}
&& D^{-1}(q_{y},\omega)=\frac{q^{z-1}_{y}}{e^{2}\Lambda^{z-3}}f\left(\frac{\omega e^{2}\Lambda^{z-3}}{q^{z}_{y}}\right), \label{eq1}\\
&& \frac{\partial e^{2}}{\partial \ln\Lambda}=(3-z)e^{2}, \label{eq2}
\end{eqnarray}
where $q_{y}$ is the momentum paralleled to the chosen momentum of the patch geometry, $z$ is the dynamical critical exponent and $e$ denotes the gauge charge with $\Lambda$ being the cutoff of the low-energy theory. The function $f(x)$ describes damping dynamics of the gauge field, its exact form depends on the details of the theory. For the case studied by Metlitski, Sachdev, Mross and Senthil,\cite{Metlitski,Mross2010} $z$ is equal to its RPA value of three, up to three-loop order and $f(x)=x$ with an extra $q^2_{y}$ term added into the gauge field propagator. In general, for a scale invariant fixed point, one could assume $f(0)=1$ and $f(x)$ approaches $x^{1-1/z}$ when $x \gg 1 $.\cite{Herbut2007} It is needed to note that the anomalous dimension $\eta$ for the gauge field is directly related to its dynamical critical exponent $z$ as $\eta=3-z$. Thus, a non-zero anomalous dimension indicates a deviation of $z$ from its RPA's result.

From the RG equation of the gauge charge $e$, one observes that for $z < 3 $ $e$ flows to an infinite coupling fixed point while for $z>3$ the situation is inverted with $e$ ending at the Gauss fixed point ($e^{\ast}=0$). Therefore, if the gauge field acquires a positive anomalous dimension (or $z<3$), the coupled system of gauge fields and Fermi surface must be strongly coupled, thus the conventional perturbative theories would not work in this situation and some nonperturbative techniques (e.g. bosonization in higher dimensions\cite{Lawler,Efetov2009,Efetov2010,Efetov2011,Mross,Zhong} and gauge/gravity duality in string theory\cite{Lee2009d,Faulkner1,Faulkner2,Liu}) are urgently desired. In contrast, the case with a zero or negative anomalous dimension ($z\geq3$) can be readily calculated by conventional perturbative techniques because the effective gauge charge, which is the expansion parameter of perturbative theories, is small in the low-energy limit as what can be seen from the RG equation.\cite{Metlitski,Mross2010} As a matter of fact, one finally recovers the expected results of RPA in this situation.\cite{Metlitski,Mross2010} Since the case of $z\geq3$ is trivial, we only focus on the gauge fields with $z<3$, namely, a possible positive anomalous dimension, in the remaining parts of the present work.

\section{dual mapping of an effective action} \label{sec3}
Having reviewed the basic scaling properties of the noncompact U(1) gauge field, we would like to utilize the scaling propagator of the noncompact gauge field as an input for an effective action which only includes degrees of freedom of gauge fields. ( One may criticize this treatment since abundant gapless fermionic excitation exists, however based on the calculation of Metlitski, Sachdev, Mross and Senthil,\cite{Metlitski,Mross2010} it seems the basic properties of the gauge field may be fully encoded in its own propagator.) The effective action for the noncompact U(1) gauge field has the following form \cite{Herbut2003,Herbut2007}
\begin{equation}
S=\frac{1}{2}\int \frac{d^{2}kd\omega}{(2\pi)^{3}}D^{-1}(k,\omega)a_{i}a_{j}\left(\delta_{ij}-\frac{k_{i}k_{j}}{k^{2}}\right), \label{eq3}
\end{equation}
where the imaginary time component of gauge fields has been dropped out due to only short-range interaction being involved (It is screened out in the low-energy limit). $a_{i}$ denotes the spatial component of the gauge field with $i=x,y$. Subsequently, by rewriting the above action in a form with explicitly gauge invariant and introducing field strength tensor $F_{ij}\equiv k_{i}a_{j}-k_{j}a_{i}$, we arrive at a lattice gauge theory in the spirit of Villain approach
\begin{eqnarray}
&& S=\sum_{x,x',\tau,\tau'}\left[\frac{1}{4}(F_{ij}(x,\tau)-2\pi n_{ij}(x,\tau))V(x-x',\tau-\tau')\right.\nonumber\\
&& \hspace{1cm}\left. \times(F_{ij}(x',\tau')-2\pi n_{ij}(x',\tau'))\right]+\frac{1}{2e^{2}}\sum_{x,\tau}[F_{i0}(x,\tau)\nonumber\\
&& \hspace{1cm} -2\pi n_{i0}(x,\tau)]^{2}, \label{eq4}
\end{eqnarray}
where the last term is a bare Maxwell term additionally and $n_{ij}$ is an integer-valued field to impose compactness of the gauge field. Thus, the above effective action indeed describes a compact U(1) gauge field theory and the dynamics of the gauge field is encoded by the effective potential $V(x,\tau)$. The effective potential $V$ in the fourier space reads $$V(k,\omega)=D^{-1}(k,\omega)k^{-2}=\frac{q^{z-3}_{y}}{e^{2}\Lambda^{z-3}}f\left(\frac{\omega e^{2}\Lambda^{z-3}}{q^{z}_{y}}\right).$$
Subsequently, it is straightforward to use the familiar Hubbard-Stratonovich transformation to decouple the quadratic term in the above lattice gauge field theory and the action obtained reads
\begin{eqnarray}
&& S=\sum_{x,x',\tau,\tau'}[c_{ij}(x,\tau)V^{-1}(x-x',\tau-\tau')c_{ij}(x',\tau')\nonumber\\
&& \hspace{1cm} +c_{i0}(x,\tau)\frac{e^{2}}{2}\delta_{xx'}\delta_{\tau\tau'}\delta_{{ij}}c_{j0}(x',\tau')]\nonumber\\
&& \hspace{1cm} +i\sum_{x,\tau}c_{\mu\nu}(x,\tau)(F_{\mu\nu}-2\pi n_{\mu\nu}). \label{eq5}
\end{eqnarray}

The integral over the gauge field can now be performed readily, and one obtains a constraint $\Delta_{\mu}c_{\mu\nu}=0$ with $\Delta_{\mu}$ being the lattice derivative. This constraint can be fulfilled by requiring $c_{\mu\nu}(x,\tau)=\frac{1}{2}\epsilon_{\mu\nu\lambda}\Delta_{\lambda}\Phi(x,\tau)$ where $\Phi(x,\tau)$ is a scalar function in the space-time and dual to monopoles. As a matter of fact, $\Phi(x,\tau)$ can be identified with the magnetic monopole potential if one integrates over it and obtains an effective action for monopoles.\cite{Herbut2007}
Then, after inserting $c_{\mu\nu}(x,\tau)=\frac{1}{2}\epsilon_{\mu\nu\lambda}\Delta_{\lambda}\Phi(x,\tau)$ into the
lattice gauge action and introducing a small fugacity $y$ in the usual way, one can obtain a sine-Gordon-like action for magnetic monopole potential $\Phi$ in the continuum limit
\begin{eqnarray}
&& S = \frac{1}{2}\int \frac{d^{2}kd\omega}{(2\pi)^{3}}[|\Phi(k,\omega)|^{2}(\omega^{2}/V(k,\omega)+e^{2}k^{2})]\nonumber\\
&& \hspace{2cm} -2y\int d^{2}xd\tau\cos(2\pi\Phi(x,\tau)). \label{eq6}
\end{eqnarray}

The physics is clear in terms of this effective action. For large fugacity $y$, the cosine potential cannot be omitted
and a mass term is obtained by magnetic monopole potential, which can be easily checked if one performs a simple variational calculation on the above sine-Gordon-like action, \cite{Nagaosa1999} thus the dual monopoles only suffer from short-range interaction mediated by the gapped magnetic monopole potential $\Phi$. Therefore, it is inevitable for monopoles to proliferate (condense) in this situation. Since the monopoles describe tunneling
events between energetically degenerated but topological inequivalent gauge vacua, the condensation of these monopoles indicates that the compactness (or periodicity) of the gauge fields is imposed firmly, thus the gauge fields have to be in the confined state. In contrast, monopoles-antimonopoles pairs are confined (bounded) due to the long-range interaction mediated by the massless (gapless) magnetic monopole potential whose masslessness results from smallness of the fugacity. In other words, when the fugacity is sufficiently small, the cosine potential can be safely neglected, which leads to gapless magnetic monopole potentials. As a result, pairs of monopoles-antimonopoles are bounded and the compactness of gauge fields are effectively relaxed, thus deconfinement of gauge fields can be fulfilled in the case of the small fugacity. To understand the details of confinement/deconfinement of gauge fields in terms of this effective action, we will turn to the RG treatment in the next section.

Before ending this section, a careful reader may doubt whether it is appropriate to use the crude sine-Gordon-like model in case of the U(1) gauge field in the presence of Fermi surface. Though a sine-Gordon-like analysis is crude and may miss some important features, for example, it cannot reflect nontrivial projective-symmetry restrictions, \cite{Hermele} but it can capture some essential physics of the confinement in the case of the U(1) gauge field in the presence of Fermi surface. \cite{Kim} Kim used the sine-Gordon-like model and correctly predicted deconfined gauge fields under an assumption that the gauge field is Landau damped with z=3 or $\eta=0$. In the present work, we follow the same way of Kim but consider $z<3$ case. When $z=3$ is taken, our result reproduce Kim's conclusion, and when $z<3$, a confined gauge field is found.

\section{renormalization group analysis of the sine-Gordon-like action} \label{sec4}
In this section, we will proceed to discuss the nature of confinement/deconfinement of gauge fields by studying the RG equations derived from the sine-Gordon-like action. To derive RG equations, the specific form of $V(k,\omega)$ or equivalently $G(k,\omega)$ must be given. However, up to now the explicit form of $G(k,\omega)$ for $\eta > 0$ (or $z < 3$) has not been reported within the framework of conventional perturbation theory up to three-loop order. Here we only consider a physically interesting limit, which corresponds to the assumption that $\omega\ll |\vec{k}|$. We note that this limit is implicitly assumed in the discussion of Landau damping of gauge fields or other order parameter fields coupled to low-energy fermions around the Fermi surface.\cite{Nagaosa1999} Therefore, if $1<z<3$, the term $\omega \gg |\vec{k}|^{z}$ is justified in the low-energy limit and one obtains a simplified form for $V(k,\omega)\approx e^{-2/z}\Lambda^{-1+3/z}|\vec{k}|^{-2}\omega^{1-1/z}$. So, inserting the simplified form of $V(k,\omega)$ into the effective action, one obtains
\begin{eqnarray}
&& S = \frac{1}{2}\int \frac{d^{2}kd\omega}{(2\pi)^{3}}[|\Phi(k,\omega)|^{2}(|\omega|^{1+\frac{1}{z}}k^{2}e^{\frac{2}{z}}\Lambda^{1-\frac{3}{z}}+e^{2}k^{2})]\nonumber\\
&& \hspace{1cm} -2y\int d^{2}xd\tau\cos(2\pi\Phi(x,\tau)), \label{eq7}
\end{eqnarray}
where at the tree level, energy $\omega$ does not rescale in the scaling transformation while momentum $k$ is rescaled to $k'=bk$ with $b$ being scaling factor. Meanwhile, the magnetic monopole potential field $\Phi(x,\tau)$ is fixed in the scaling transformation since it must be a non-dimensional quantity appearing in the cosine potential.

Having these preliminary results in hand, it is straightforward to derive iterative relations of the gauge charge $e^{2}$ and the fugacity $y$ up to the second order of the fugacity as follows
\begin{eqnarray}
&& y(b)=yb^{2}e^{-\frac{1}{2}G_{>}(0,0)}, \label{eq8}\\
&& e^{2}(b)=e^{2}+\frac{1}{2}y^{2}e^{-G_{>}(0,0)}\int d^{2}xd\tau x^{2}(e^{G_{>}(x,\tau)}-1). \label{eq9}
\end{eqnarray}
 Here we have defined an auxiliary function $G_{>}(x,\tau)=(2\pi)^{2}\int_{\Lambda/b<|\vec{k}|<\Lambda}\frac{d^{2}k}{(2\pi)^{2}}\int \frac{d\omega}{2\pi}e^{i(\vec{k}\cdot\vec{x}-\omega\tau)}\frac{1}{k^{2}}\frac{1}{e^{2}+e^{2/z}|\omega|^{1+\frac{1}{z}}\Lambda^{1-\frac{3}{z}}}$.
The corresponding RG equations read
\begin{eqnarray}
&& \frac{dy}{d\ln b}=\left(2-\frac{C}{2e^{4/(1+z)}}\right)y, \label{eq10}\\
&& \frac{de^{2}}{d\ln b}=\frac{(2\pi)^{2}A}{e^{2}}y^{2},\label{eq11}
\end{eqnarray}
where $A,C$ are positive nonuniversal numerical constants whose precise values are not important in the present consideration. For a detailed derivations of these two equations, one can refer to Appendixes A and B. Then, following the argument of Kim\cite{Kim} on the deconfinement of compact U(1) gauge field coupled to Fermi surface, we can combine the RG equation proposed by Metlitski and Sachdev\cite{Metlitski} with the above equations from the effective sine-Gordon-like action. The resulting RG equations read
\begin{eqnarray}
&& \frac{dy}{d\ln b}=\left(2-\frac{C}{2e^{4/(1+z)}}\right)y, \label{eq12}\\
&& \frac{de^{2}}{d\ln b}=(3-z)e^{2}+\frac{(2\pi)^{2}A}{e^{2}}y^{2}.\label{eq13}
\end{eqnarray}

It is noted that if we consider the case of $z=3$, the above RG equations will recover the familiar results of conventional sine-Gordon action in d=2 except for $e^{4/(1+z)}$ in the denominator of the first RG equation.\cite{Herbut2007} For further discussion, it is useful to use the electromagnetic duality $g=e^{-2}$ where $g$ is the magnetic charge carried by monopoles(Actually, the real magnetic charge is $q=\sqrt{g}$, but for simplicity we use $g$ as magnetic charge in what follows.). Thus, we get the following RG equations
\begin{eqnarray}
&&\frac{dy}{d\ln b}=\left(2-\frac{Cg^{\frac{2}{1+z}}}{2}\right)y, \label{eq14}\\
&&\frac{dg}{d\ln b}=-(3-z)g-(2\pi)^{2}Ag^{3}y^{2}.\label{eq15}
\end{eqnarray}
The fixed points can be readily found by utilizing these two RG equations and we have a solution of fixed points $g^{\ast}=0,y^{\ast}=0$ for the case of $z<3$ (or $\eta>0$). One can see that this fixed point is unstable thus the RG flow will end at $g^{\ast}=0,y^{\ast}=\infty$. As a result, the fugacity of monopoles flows to infinite fixed point and monopoles are condensing in this condition, which leads to a confined state for U(1) gauge fields in presence of Fermi surface and the putative quantum spin liquid unstable to other possible symmetry-breaking states, e.g., valence bond solids and N\'eel antiferromagnet. The nature of physics of the above analysis is that when the gauge field gains a positive anomalous dimension from its strongly coupling to gapless fermions around Fermi surface, the quasiparticle-like feature is lost, thus the gauge field itself is fluctuating rather strongly. Therefore, the compactness of gauge fields has to be taken into account firmly, which means the gauge fields must be in the confined state. This is in contrast to the case of weakly fluctuating, in which the compactness can be effectively relaxed and the deconfinement of gauge fields is obtained.

Recently, some attempts like the derivative of strings theory, the anti-de Sitter/conformal field theory (AdS/CFT) correspondence (or gauge/gravity duality), have been used to attack the perplexing non-Fermi liquid behaviors in strongly correlated electronic systems \cite{Lee2009d,Faulkner1,Faulkner2,Liu} and a clear correspondence is found between certain quantum gravity theories with an emergent AdS$_{2}\times R^{d-1}$ geometry and fractionalized metallic states of Kondo-Heisenberg model. \cite{Sachdev3} To our knowledge, in the presence of Fermi surface no positive anomalous dimensions for U(1) gauge fields have been reported in the literature. Therefore, it is interesting to see whether the AdS/CFT correspondence could provide a first example with nontrivial positive anomalous dimension for gauge fields in the presence of Fermi surface in the near future. If so, the result here will have a chance to communicate with the string theories and could be inspected from the viewpoint of quantum gravity.

\section{Conclusion}\label{sec5}
In the present paper we have shown that\textit{if the U(1) gauge field has a possible positive anomalous dimension in the presence of Fermi surface}, it will have to be in the confined state due to strong fluctuations induced by gapless fermionic excitations on the Fermi surface in two spatial dimensions. Therefore it indicates that some quantum spin liquids and/or fractionalized metallic states described in terms of this kind of gauge fields are unstable. The corresponding fractionalized degrees of freedom have to be recombined into conventional elementary excitations such as electrons and magnons in related systems. However, to our knowledge, a positive anomalous dimensions for U(1) gauge fields have not been reported in the presence of Fermi surface in the literature. We hope that the result could be useful for further study in the quantum spin liquids and many related slave-particle gauge theories in strongly correlated electron systems. Probably, the flourish AdS/CFT correspondence could provide an example we want and our result may be inspected with the viewpoint of quantum gravity in the near future.

\begin{acknowledgments}
The authors would like to thank Igor Herbut for useful discussion and suggestion on the relation to $cQED_{3}$ and Ki-Seok Kim for helpful correspondence. The work is supported partially by the Program for NCET, NSF and the Fundamental Research Funds for the Central Universities of China.
\end{acknowledgments}

\appendix

\section{Derivation for the RG equation of the fugacity $y$}
Here we give the details how one can derive the RG equation of the fugacity $y$ [Eq. (\ref{eq10})] from Eq. (\ref{eq8}), which reads $y(b)=yb^{2}e^{-\frac{1}{2}G_{>}(0,0)}$
and
\begin{eqnarray}
&& G_{>}(0,0) = (2\pi)^{2}\int_{\Lambda/b<|\vec{k}|<\Lambda}\frac{d^{2}k}{(2\pi)^{2}} \nonumber\\
&&\times \int\frac{d\omega}{2\pi}\frac{1}{k^{2}}\frac{1}{e^{2}+e^{\frac{2}{z}}|\omega|^{1+\frac{1}{z}}\Lambda^{1-\frac{3}{z}}}\nonumber\\
&&=\int_{\Lambda/b}^{\Lambda}\frac{dk}{k}\int_{-\infty}^{\infty}d\omega\frac{1}{e^{\frac{2}{z}}/\Lambda^{\frac{3}{z}-1}}\nonumber\\
&&\times\frac{1}{e^{2-\frac{2}{z}}\Lambda^{\frac{3}{z}-1}+|\omega|^{1+\frac{1}{z}}}\nonumber\\
&&=\frac{2\Lambda^{\frac{3}{z}-1}}{e^{\frac{2}{z}}}\ln b\int_{0}^{\infty}\frac{1}{e^{2-\frac{2}{z}}\Lambda^{\frac{3}{z}-1}+|\omega|^{1+\frac{1}{z}}}\nonumber\\
&&=\frac{2\Lambda^{\frac{3}{z}-1}}{e^{\frac{2}{z}}}\ln b \frac{\pi(e^{2-\frac{2}{z}}\Lambda^{\frac{3}{z}-1})^{-\frac{1}{z+1}}}{(\frac{z+1}{z})\sin \frac{\pi z}{z+1}}\nonumber
\end{eqnarray}
So the $G_{>}(0,0)$ can be written as $G_{>}(0,0)=\frac{C\ln b}{e^{\frac{4}{z+1}}}$ with $C$ being a positive constant and the resulting scale transformation of the fugacity $y(b)$ reads $y(b)=b^{2-\frac{C}{2e^{4/(z+1)}}}y$. Therefore, the RG equation can be readily find as $\frac{dy}{d\ln b}=\left(2-\frac{C}{2e^{4/(1+z)}}\right)y$, namely, Eq.(\ref{eq10}).

\section{Derivation for the RG equation of the gauge charge $e^{2}$}
In this section, we will derive the RG equation of $e^{2}$[Eq. (\ref{eq11})] from Eq. (\ref{eq9}).
For derivation of the RG equation of $e^{2}$, we first have to calculate
$G_{>}(x,\tau)=(2\pi)^{2}\int_{\Lambda/b<|\vec{k}|<\Lambda}\frac{d^{2}k}{(2\pi)^{2}}\int \frac{d\omega}{2\pi}e^{i(\vec{k}\cdot\vec{x}-\omega\tau)}\frac{1}{k^{2}}\frac{1}{e^{2}+e^{2/z}|\omega|^{1+\frac{1}{z}}\Lambda^{1-\frac{3}{z}}}$.
According to the treatment of Ref. \onlinecite{Herbut2007}, the $G_{>}(x,\tau)$ can be found as $G_{>}(x,\tau)=\int_{\Lambda/b}^{\Lambda}\frac{dkJ_{0}(kx)}{k}\int d\omega\frac{e^{-i\omega\tau}}{e^{2-\frac{2}{z}}\Lambda^{\frac{3}{z}-1}+|\omega|^{1+\frac{1}{z}}}$.
where $J_{0}(kx)$ is the Bessel function. Then using $\int_{\Lambda/b}^{\Lambda}\frac{dkJ_{0}(kx)}{k}=-x\Lambda \ln b \frac{dK_{0}(z)}{dz}|_{z=x\Lambda}=x\Lambda\ln b K_{1}(x\Lambda)$ with $K_{0},K_{1}$ being the modified Bessel function.
\begin{eqnarray}
&&\int d^{2}xd\tau x^{2}(e^{G_{>}(x,\tau)}-1)= \nonumber\\
&&\frac{2\pi}{\Lambda^{4}}\int_{0}^{\infty}dz z^{4}K_{1}(z)\frac{\Lambda^{\frac{3}{z}-1}}{e^{\frac{2}{z}}}\int d\tau d\omega\frac{e^{-i\omega\tau}}{e^{2-\frac{2}{z}}\Lambda^{\frac{3}{z}-1}+|\omega|^{1+\frac{1}{z}}}\nonumber\\
&&=\frac{32\pi}{\Lambda^{4}}\ln b\frac{\Lambda^{\frac{3}{z}-1}}{e^{\frac{2}{z}}}\int d\tau d\omega\frac{e^{-i\omega\tau}}{e^{2-\frac{2}{z}}\Lambda^{\frac{3}{z}-1}+|\omega|^{1+\frac{1}{z}}}\nonumber\\
&&=\frac{64\pi^{2}}{e^{2}\Lambda^{4}}\ln b=\frac{A}{e^{2}}\ln b. \nonumber
\end{eqnarray}
Therefore, we obtain $e^{2}(b)=e^{2}+\frac{(2\pi)^{2}Ay^{2}}{e^{2}}\ln b$ and the corresponding RG equation is
$\frac{de^{2}}{d\ln b}=(3-z)e^{2}+\frac{(2\pi)^{2}A}{e^{2}}y^{2}$, i.e., Eq. (\ref{eq11}).

\end{document}